\documentclass[epj]{webofc}
\usepackage[varg]{txfonts}   
\woctitle{Nobel Symposium NS160: Chemistry and Physics of Heavy and Superheavy Elements}
\begin{document}
\title{Characterizing the mechanism(s) of heavy element synthesis\\ reactions}
%
%

\author{\firstname{Walter} \lastname{Loveland}\inst{1}\fnsep\thanks{\email{lovelanw@onid.orst.edu}} 
}

\institute{Department of Chemistry, Oregon State University, Corvallis, 
OR 97331, USA
          }

\abstract{%
  A review of the current state of our understanding of  complete 
fusion  reaction mechanisms is presented, 
from the perspective of an experimentalist.  For complete fusion reactions, 
the overall uncertainties in predicting heavy element synthesis cross 
sections are examined in terms of the uncertainties associated with the 
calculations of capture cross sections, fusion probabilities and survival 
probabilities. 
}
\maketitle
\section{Introduction}
\label{intro}
\begin{figure}[th]
\vspace*{-6mm}
\centering
\includegraphics[width=77mm]{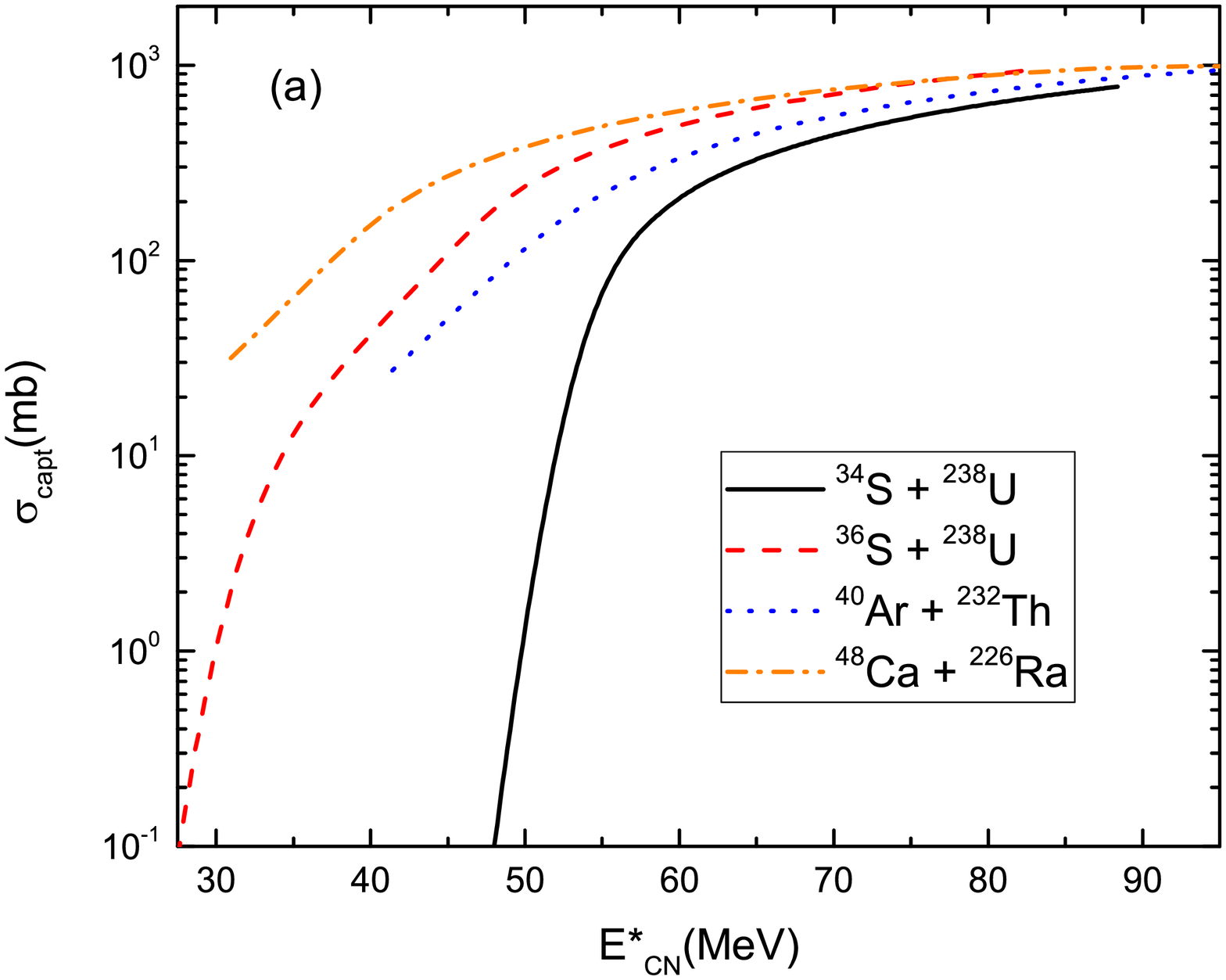}
\hspace*{-10mm}
\includegraphics[width=73mm]{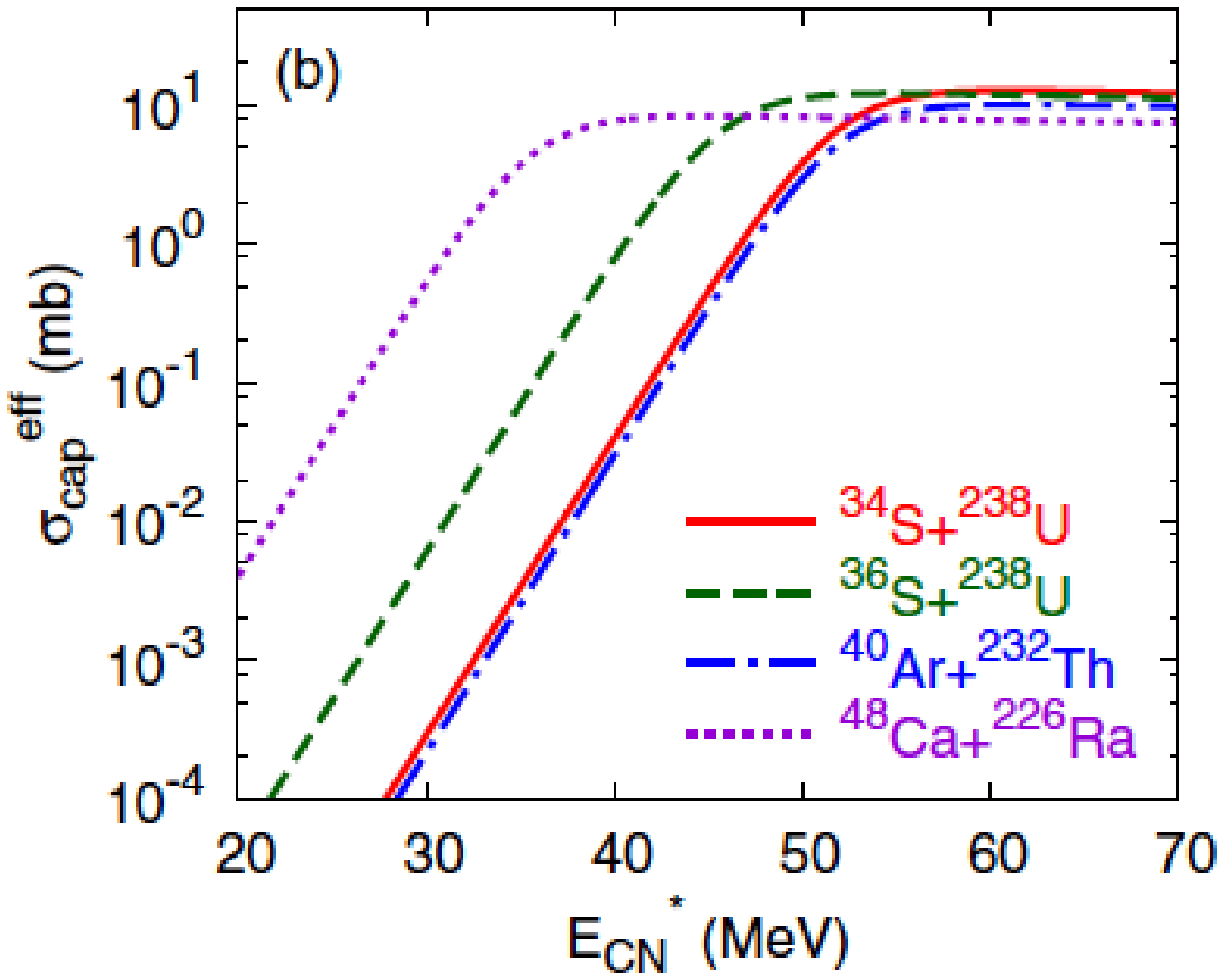}
\vspace*{-7mm}
\caption{(a) Calculated capture cross sections for some typical reactions.
(b) the ``spin-limited'' capture cross sections for the reactions in (a) 
\cite{r1}.}
\label{WL-f1}       
\end{figure}

Formally, the cross section for producing a heavy evaporation residue, 
$\sigma_{\rm EVR}$, in a fusion reaction can be written as
\begin{equation}
\sigma_{\rm EVR}(E)=\frac{\pi h^2}{2\mu E}\sum\limits_{\ell=0}^\infty
(2\ell+1)T(E,\ell)P_{\rm CN}(E,\ell)W_{\rm sur}(E,\ell)
\end{equation}
where $E$ is the center of mass energy, and $T$ is the probability of the 
colliding nuclei to overcome the potential barrier in the entrance channel 
and reach the contact point. (The term ``evaporation residue'' refers to the 
product of a fusion reaction followed by the evaporation of a specific 
number of neutrons.) $P_{\rm CN}$ is the probability that the 
projectile-target system will evolve from the contact point to the 
compound nucleus. $W_{\rm sur}$ is the probability that the compound 
nucleus will decay to produce an evaporation residue rather than fissioning.  
Conventionally the EVR cross section is separated into three individual
reaction stages (capture, fusion, survival) motivated, in part, by the 
different time scales of the processes. However, one must remember that the 
$W_{\rm sur}$ term effectively sets the allowed values of the spin. This 
effect is shown in figure~\ref{WL-f1} where the capture cross sections for 
several reactions are shown without and with the spin limitation posed by the 
survival probabilities. 

\begin{figure}[t]
\vspace*{0mm}
\centering
\includegraphics[width=75mm]{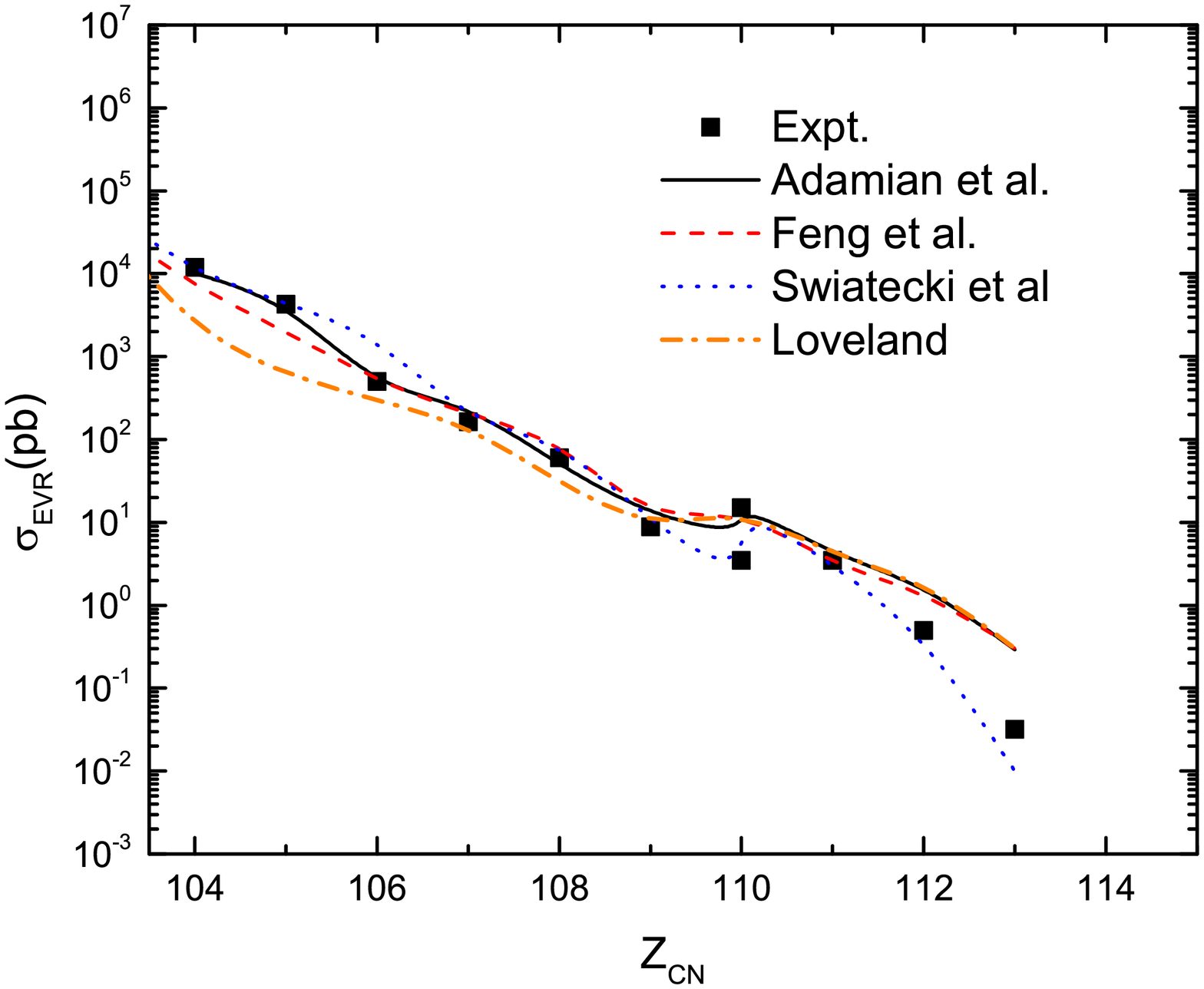}
\hspace*{-10mm}
\includegraphics[width=75mm]{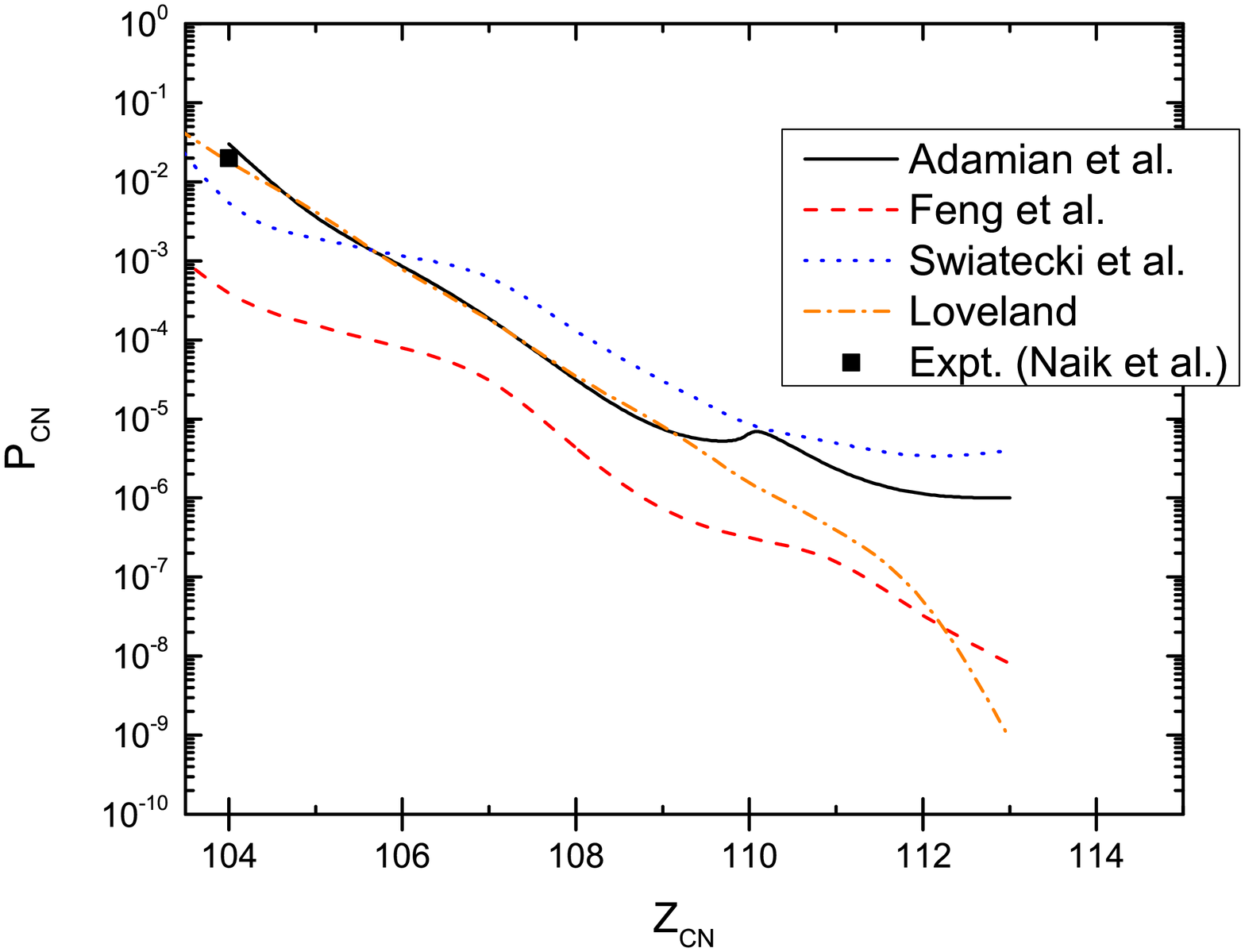}
\unitlength1mm
\begin{picture}(0,0)
\put(-55,50){(a)}
\put(15,50){(b)}
\end{picture}
\vspace*{-6mm}
\caption{(a) Typical predictions of the formation cross sections of 
elements 104-113 using cold fusion reactions. 
(b) $P_{\rm CN}$ values for the predictions in panel (a). 
The references cited in the legends refer to Adamian \cite{r4}, 
Feng \cite{r5}, Swiatecki \cite{r2}, and Loveland \cite{r3}. 
The additional reference in panel (b) is to the data of Naik 
{\it et al.} \cite{r6}.}
\label{WL-f2}       
\end{figure}

Several successful attempts have been made to describe the cross sections 
for evaporation residue formation in cold fusion reactions \cite{r2,r3,r4,r5}. 
In figure~\ref{WL-f2}(a), I show some typical examples of post dictions of the 
formation cross sections for elements 104-113 in cold fusion reactions. 
The agreement between theory and experiment is impressive because the cross 
sections extend over six orders of magnitude, i.e., a robust agreement. 
Because the values of $\sigma_{\rm capture}$ are well known or generally 
agreed upon (see below), the values of the product 
$P_{\rm CN}\cdot W_{\rm sur}$ are the same in most of these post dictions.  
However, as seen in figure~\ref{WL-f2}(b), 
the values of $P_{\rm CN}$ differ significantly in these 
post dictions \cite{r2,r3,r4,r5}, and differ from measurements of $P_{\rm CN}$
\cite{r6}. A similar situation occurs in predictions of cross sections for 
hot fusion reactions. These are clear-cut cases 
in which a simple agreement between theory and experiment in postdicted 
cross sections is not sufficient to indicate a real understanding of the 
phenomena involved.

We might ask what the overall uncertainties are in the current 
phenomenological models for predicting heavy element production cross 
sections. This is an item of some controversy. Some feel the uncertainties 
in typical predictions are factors of 2-4 \cite{r7} while others 
estimate these uncertainties to be 1-2 orders of magnitude \cite{r8,r9}.

\section{Capture Cross Sections}
\label{sec2}
The capture cross section is, in the language of coupled channel calculations, 
the ``barrier crossing'' cross section. It is the sum of the quasifission, 
fast fission, fusion-fission and fusion-evaporation residue cross sections. 
The barriers involved are the interaction barriers and not the fusion 
barriers. There are several models for capture cross sections. Each of them 
has been tested against a number of measurements of capture cross sections 
for reactions that, mostly, do not lead to the formation of the heaviest 
nuclei. In general, these models are able to describe the magnitudes of the 
capture cross sections within 50\% and the values of the interaction barriers 
within 20\%. The most robust of these models takes into account the effects 
of target and projectile orientation/deformation upon the collisions, the 
couplings associated with inelastic excitations of the target and projectile 
and the possibility of one or two neutron transfer processes. Loveland 
\cite{r10} has compared calculations of the capture cross sections for 
reactions that synthesize heavy nuclei with the measured cross sections. 
Good agreement between the measured and calculated values of the cross 
sections occurs for all reactions. The ratio of calculated to observed 
capture cross sections varies from 0.5 to 2. Nominally, given the other 
uncertainties in estimating $\sigma_{\rm EVR}$, this seems generally 
acceptable. However, from the point of view of an experimentalist, it is 
not acceptable. The capture cross section is relatively easy to measure 
and an uncertainty of a factor of 50\% may mean having to run an experiment 
for several months longer to be successful in a synthetic effort. 

\begin{figure}[b]
\centering
\includegraphics[width=75mm]{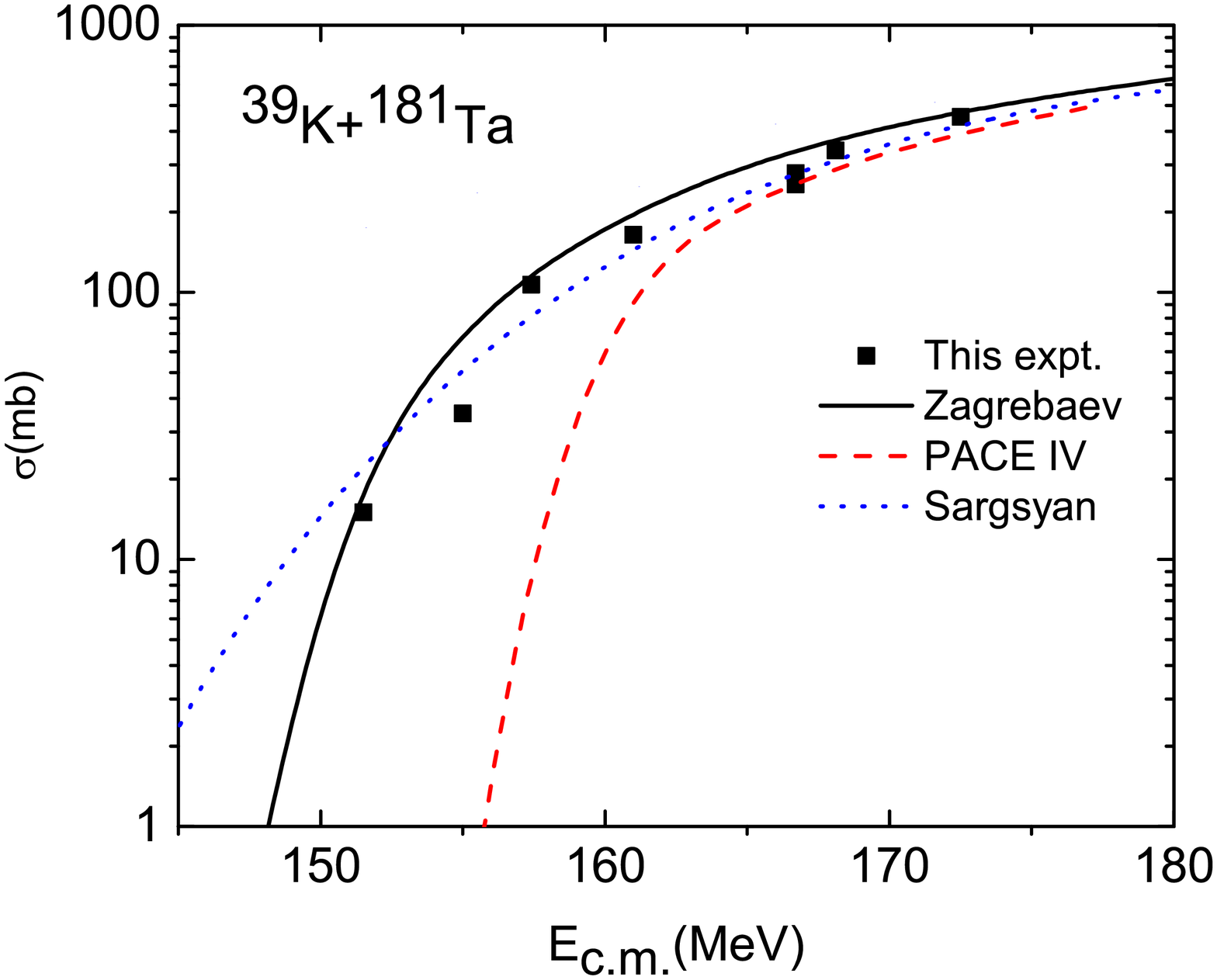}
\hspace*{-10mm}
\includegraphics[width=75mm]{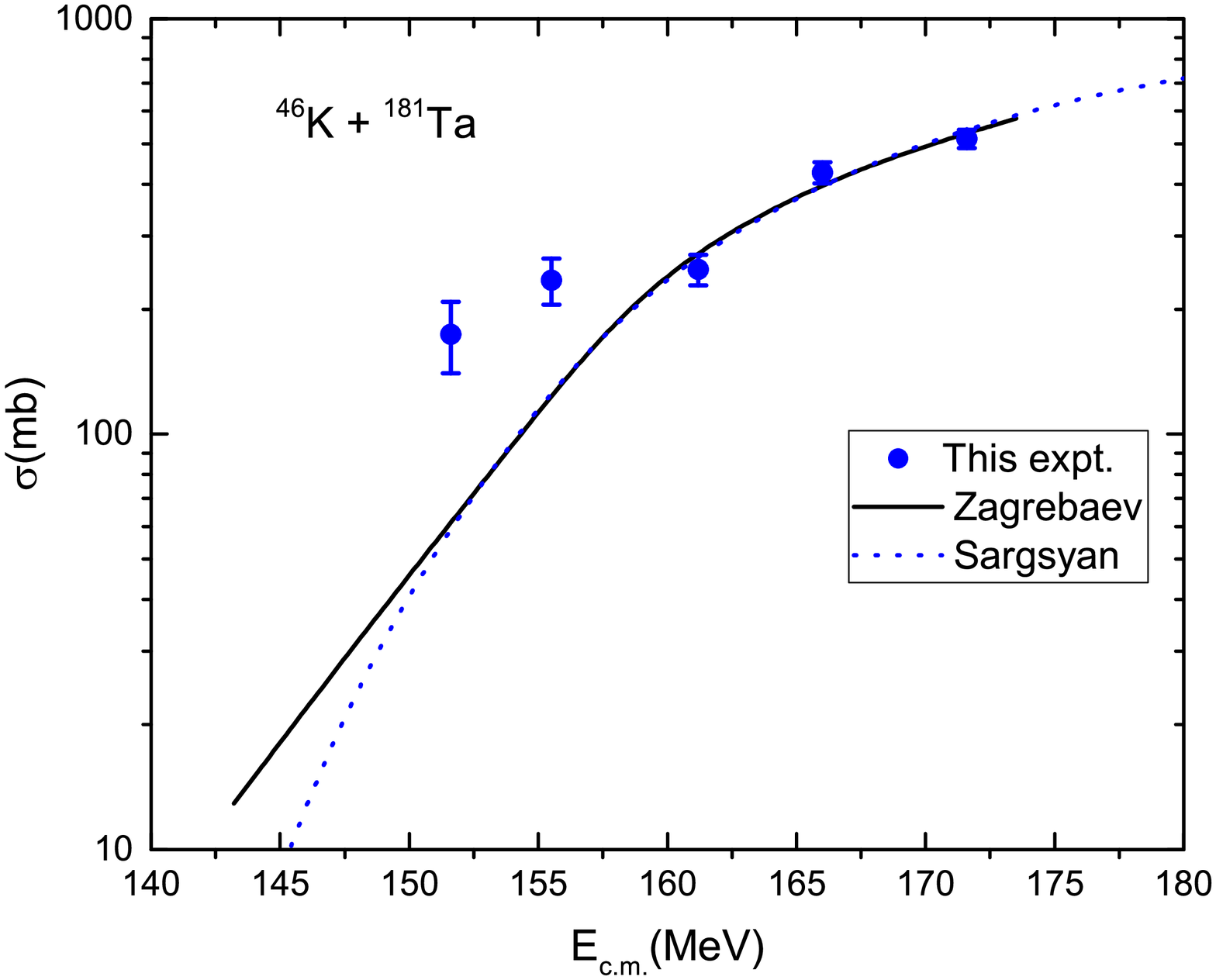}
\vspace*{-6mm}
\caption{The capture-fission excitation functions for the reaction 
of $^{39}$K (left) and $^{46}$K (right) with $^{181}$Ta.}
\label{WL-f4}       
\end{figure}

Future synthetic efforts with heavy nuclei may involve the use of very 
neutron-rich beams such as radioactive beams. While we understand that 
such efforts are not likely to produce new superheavy elements due to the 
low intensities of the radioactive beams \cite{r3}, there may exist a window 
of opportunity to make new neutron-rich isotopes of elements 104-108 
\cite{r11}. Our ability to predict the capture cross sections for the 
interaction of very neutron-rich projectiles with heavy nuclei is limited 
\cite{r11} and this is especially true near the interaction barrier 
where the predictions of models of the capture cross sections may differ 
by orders of magnitude \cite{r11}. As part of the effort to study the 
scientific issues that will be relevant at next generation radioactive 
beam facilities, such as FRIB, we have started to use the ReA3 facility at 
the NSCL to study capture processes with fast-releasing beams such as 
the potassium isotopes.  We chose to study the interaction of 
$^{39,46}$K with $^{181}$Ta.  The first {\em preliminary} results 
from that experiment are shown in figure~\ref{WL-f4}.  The $^{39}$K +$^{181}$Ta 
reaction results seem to be well understood within conventional pictures 
of capture \cite{r12,r13} while the neutron-rich $^{46}$K results suggest 
an unusual near barrier fusion enhancement.

\section{Survival Probabilities, $W_{\rm sur}$}
\label{sec3}
Formally $W_{\rm sur}$ can be written as
\begin{equation}
W_{\rm sur}(E_{\rm c.m.})=P_{xn}(E^*_{\rm CN})\prod\limits_{i=1}^x 
\frac{\mathit{\Gamma}_n(E^*_i)}{\mathit{\Gamma}_n(E^*_i)+
\mathit{\Gamma}_f(E^*_i)}
\end{equation}
where $P_{xn}$ is the probability of emitting $x$ (and only $x$) neutrons 
from a nucleus with excitation energy $E^*$, and $\mathit{\Gamma}_n$ and $\mathit{\Gamma}_f$ 
are the partial widths for decay of the completely fused system by either 
neutron emission or fission, respectively. For the most part, the formalism 
for calculating the survival, against fission, of a highly excited nucleus 
is understood. There are significant uncertainties, however, in the input 
parameters for these calculations and care must be used in treating some 
situations.  ``Kramers effects'' and the overall fission barrier height 
are found \cite{r14} to have the biggest effect on the calculated cross 
sections.

A recent experiment concerning survival probabilities in hot fusion 
reactions showed the importance of ``Kramers effects'' \cite{r15}. The 
nucleus $^{274}$Hs was formed at an excitation energy of 63~MeV using the 
$^{26}$Mg+$^{248}$Cm reaction. $^{274}$Hs has several interesting properties. 
The liquid drop model fission barrier height is zero and there is a 
subshell at $N = 162$, $Z = 108$. In the formation reaction, $P_{\rm CN}$ 
is measured \cite{r16} to be 1.0. By measuring the angular distribution 
of the fission associated neutrons, Yanez {\it et al.} \cite{r15} were able 
to deduce a value of $\mathit{\Gamma}_n/\mathit{\Gamma}_{\rm total}$ for 
the first chance 
fission of $^{274}$Hs ($E^* = 63$~MeV) of $0.89\pm 0.13$! A highly excited 
fragile nucleus with a vanishingly small fission barrier decayed $\sim 90$\% 
of the time by emitting a neutron rather than fissioning. Conventional 
calculations with various values of the fission barrier height were unable 
to reproduce these results. The answer to this dilemma is to consider the 
effects of nuclear viscosity to retard fission \cite{r17}, the so-called 
Kramers effects. These Kramers effects are the reason that hot fusion 
reactions are useful in heavy element synthesis, in that the initial high 
excitation energies of the completely fused nuclei do not result in 
catastrophic losses of surviving nuclei \cite{r18}.

With respect to fission barrier heights, most modern models do equally 
well/poorly in describing fission barrier heights for Th-Cf nuclei. 
Afanasjev {\it et al.} \cite{r19} found the average deviation between 
the calculated and known inner barrier heights was $\sim 0.7$~MeV amongst 
various models. Bertsch {\it et al.} \cite{r20} estimate the uncertainties 
in fission barrier heights are 0.5-1.0~MeV in known nuclei. Kowal {\it et al.}
\cite{r21} found for even-even nuclei with $Z = 92$-$98$ the difference 
between measured and calculated inner barrier heights was 0.8~MeV. 
Baran {\it et al.} \cite{r22} found very large, i.e., several MeV, 
differences between various calculated fission barrier heights for 
$Z = 112$-$120$. In summary, fission barrier heights are known within 
0.5-1.0~MeV. For super-heavy nuclei, the change of fission barrier height 
by 1~MeV in each neutron evaporation step can cause an order of magnitude 
uncertainty in the $4n$-channel. For the $3n$-channel, the uncertainty 
is about a factor of four \cite{r14}. 

An additional problem is that at the high excitation energies characteristic 
of hot fusion reactions, the shell effects stabilizing the fission barrier 
are predicted \cite{r23} to be ``washed out'' with a resulting fission 
barrier height $< 1$~MeV for some cases. Furthermore the rate of damping 
of the shell effects differs from nucleus to nucleus. This point is well 
illustrated by the calculations in reference~\cite{r24} of the ``effective'' 
fission barrier heights for the $^{48}$Ca+$^{249}$Bk reaction. 

Measurements of fission barrier heights are difficult and the results 
depend on the models used in the data analysis.  Recently Hofmann {\it et al.}
\cite{r25} have deduced the shell-correction energies from the systematics 
of the $Q_\alpha$ values for the heaviest nuclei and used these 
shell-correction energies to deduce fission barrier heights.  The deduced 
barrier heights for elements 118 and 120 may be larger than expected. 

\section{Fusion Probability, $P_{\rm CN}$}
\label{sec4}
The fusion probability, $P_{\rm CN}$, is the least known (experimentally) 
of the factors affecting complete fusion reactions and perhaps the most 
difficult to model.  The essential task is to measure the relative amounts 
of fusion-fission and quasifission in a given reaction.  Experimentally this 
is done using mass-angle correlations where it is difficult to measure, 
with any certainty, the fraction of fusion reactions when that quantity is 
less than 1\%. (For cold fusion reactions, $P_{\rm CN}$ is predicted to 
take on values of $10^{-2}$ to $10^{-6}$ for reactions that make elements 
104-113.) The reaction of $^{124}$Sn with $^{96}$Zr can be used to illustrate 
the uncertainties in the theory to estimate $P_{\rm CN}$ where 
\cite{r10} various theoretical estimates of $P_{\rm CN}$ range from 
0.0002 to 0.56 and where the measured value is 0.05 as well as the 
data shown in figure~\ref{WL-f2}(b).  

Where we have made progress in understanding $P_{\rm CN}$ is in the 
excitation energy dependence of $P_{\rm CN}$. Zagrebaev and Greiner
\cite{r26} have suggested the following ad hoc functional form for the 
excitation energy dependence of $P_{\rm CN}$ 
\begin{equation}
P_{\rm CN}(E^*,J)=\frac{P^0_{\rm CN}}{1+\mbox{exp}
[\frac{E^*_B-E^*_{\rm int}(J)}{\Delta}]}
\end{equation}
where $P_{\rm CN}^0$ is the fissility-dependent ``asymptotic'' (above 
barrier) value of $P_{\rm CN}$ at high excitation energies, $E_B^*$ is 
the excitation energy at the Bass barrier, $E_{\rm int}^*(J)$ is the internal 
excitation energy ($E_{c.m.}+Q-E_{\rm rot}(J)$), $J$ is the angular momentum 
of the compound nucleus, and $\Delta$ (an adjustable parameter) is taken to 
be 4~MeV.  This formula describes the extensive data of Knyazheva {\it et al.}
\cite{r27} for the $^{48}$Ca+$^{154}$Sm reaction very well \cite{r10}.  
A generalization of this formula has been used to describe the excitation 
energy dependence of $P_{\rm CN}$ for the reactions of $^{48}$Ca with 
$^{238}$U, $^{244}$Pu and $^{248}$Cm \cite{r28}.

\begin{figure}[th]
\centering
\includegraphics[width=78mm]{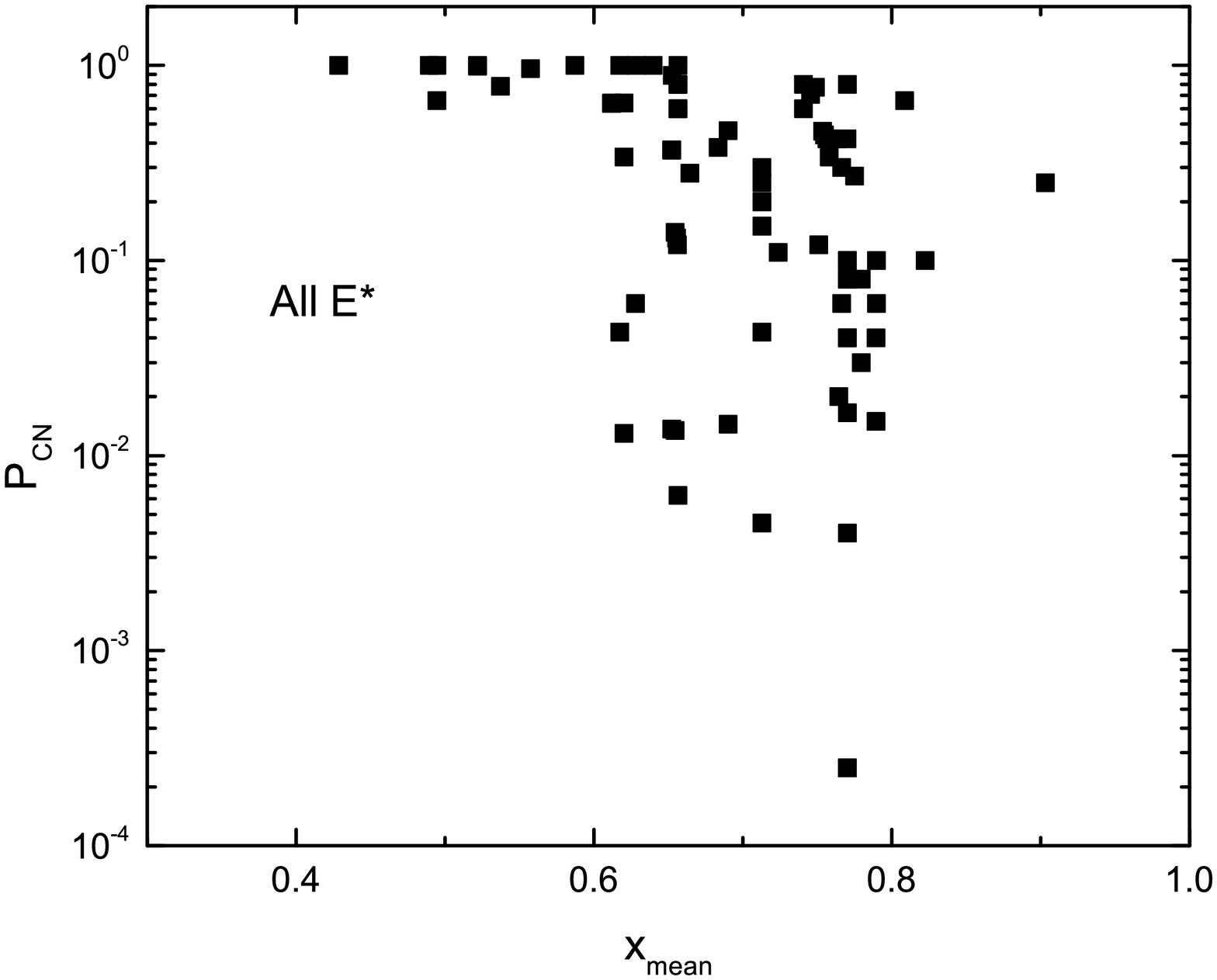}
\hspace*{-16mm}
\includegraphics[width=78mm]{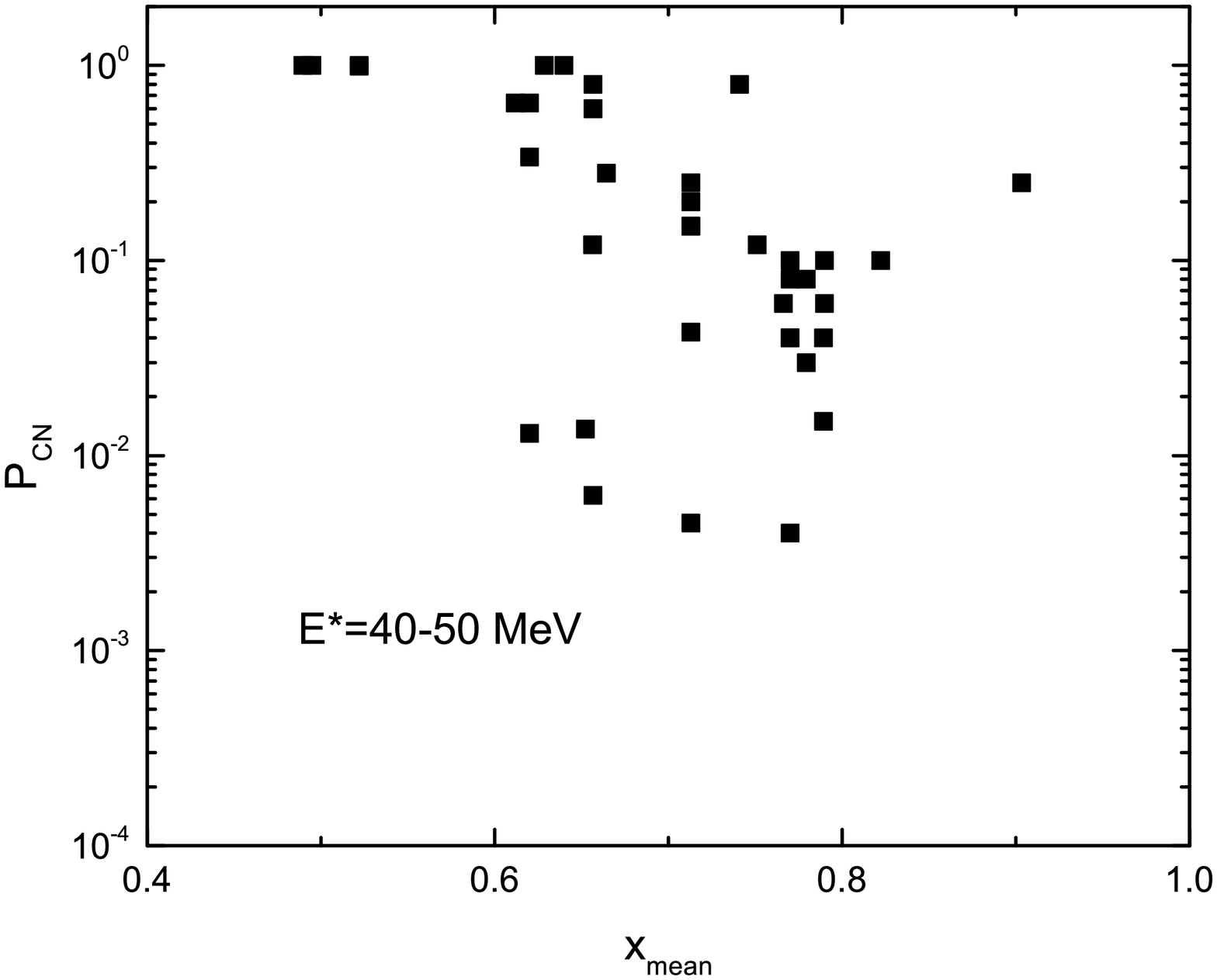}
\vspace*{-8mm}
\caption{Fissility dependence of $P_{\rm CN}$ with and without 
excitation energy sorting.}
\label{WL-f5}       
\end{figure}

It is also clear that $P_{\rm CN}$ must depend on the entrance channel 
asymmetry of the reaction. Numerous scaling factors to express this 
dependence have been proposed and used. An extensive survey of $P_{\rm CN}$
 in a large number of fusing systems was made by du~Rietz {\it et al.} 
\cite{r29}. They thought that perhaps some fissility related parameter 
would be the best way to organize their data on $P_{\rm CN}$ and its 
dependence of the properties of the entrance channel in the reactions 
they studied. They found the best fissility-related scaling variable 
that organized their data was $x_{\rm du Rietz} = 0.75x_{\rm eff} + 
0.25x_{\rm CN}$. The parameters $x_{\rm eff}$ and $x_{\rm CN}$ are the 
associated fissilities for the entrance channel and the compound system, 
respectively. The following equations can be used to calculate these 
quantities
\begin{displaymath}
x_{\rm CN}=\frac{(Z^2/A)_{\rm CN}}{(Z^2/A)_{\rm critical}};
(Z^2/A)_{\rm critical}=50.883\cdot[1.-1.7826(\frac{A-2Z}{A})^2];
x_{\rm eff}=\frac{4Z_1Z_2/[A_1^{1/3}A_2^{1/3}(A_1^{1/3}+A_2^{1/3})]}{(Z^2/A)_{\rm critical}}
\end{displaymath}
In figure~\ref{WL-f5}, I show most of the known data on $P_{\rm CN}$ using 
the du~Rietz scaling variable.  There is no discernable pattern.  
Restricting the choice of cases to those in a narrow excitation energy 
bin improves the situation somewhat but it is clear we are missing 
something in our semi-empirical systematics.  
	
Some progress has been made in calculating $P_{\rm CN}$ using TDHF 
calculations. Wakhle {\it et al.} \cite{r30} made a pioneering study of 
the $^{40}$Ca+$^{238}$U reaction. The capture cross sections predicted by 
their TDHF calculations agreed with measured capture cross sections 
\cite{r31} within $\pm 20$\%. In addition they were able to predict the 
ratio of fusion to capture cross sections, 
$\sigma_{\rm fus}/\sigma_{\rm capture}$, as $0.09\pm 0.07$ at 205.9~MeV and 
$0.16\pm 0.06$ at 225.4~MeV in agreement with reference~\cite{r31} who 
measured ratios of $0.06\pm 0.03$ and $0.14\pm 0.05$, respectively. Whether 
TDHF calculations will become a predictive tool for heavy element synthesis 
remains to be seen.

\section{Predictions for the Production of Elements 119 and 120}
\label{sec5}
Loveland \cite{r10} has shown that the current predictions for the 
production cross sections for elements 119 and 120 differ by 1-3 orders 
of magnitude, reflecting the uncertainties discussed above.  
For the reaction $^{50}$Ti+$^{249}$Bk$\rightarrow 119$, the uncertainties 
in the predicted maximum cross sections for the $3n$ and $4n$ channels 
differ by ``only'' a factor of 20-40 while larger uncertainties are found 
in the predictions for the $^{54}$Cr+$^{248}$Cm$\rightarrow 120$ reaction.  
The energies of the maxima of the $3n$ and $4n$ excitation functions are 
uncertain to 3-4~MeV, a troublesome situation for the experimentalist.

\section{Conclusions}
\label{sec6}
I conclude that: (a) Capture cross sections should be measured for reactions 
of interest. (b) We need better and more information on fission barrier 
heights, and their changes with excitation energy for the heaviest nuclei.
(c) We need to devise better methods of measuring $P_{\rm CN}$ and more 
TDHF calculations of $P_{\rm CN}$. (d) The current uncertainty in 
calculated values of $\sigma_{\rm EVR}$ is at least 1-2 orders of magnitude. 
(e) New opportunities for making neutron-rich actinides with RNBs may exist.\\

\begin{acknowledgement}
This work was supported, in part, by the Director, Office of Energy Research, 
Division of Nuclear Physics of the Office of High Energy and Nuclear 
Physics of the U.S.~Department of Energy under Grant DE-SC0014380 and 
the National Science Foundation under award 1505043.
\end{acknowledgement}


\begin{thebibliography}{99}
%
%
\bibitem{r1}
J.~Hong, G.G.~Adamian,  N.V.~Antonenko, Phys.~Rev.~C~{\bf 92}, 014617 (2015).

\bibitem{r2}
W.J.~Swiatecki, K.~Siwek-Wilczynska, J.~Wilczynski, Phys.Rev.~C~{\bf 71}, 
014602 (2005).

\bibitem{r3}
W.~Loveland, Phys.~Rev.~C~{\bf 76}, 014612 (2007).

\bibitem{r4}
G.G.~Adamian, N.V.~Antonenko, W.~Scheid, Nucl.~Phys.~{\bf A678}, 24 (2000).

\bibitem{r5}
Z.-Q.~Feng, G.-M.~Jin, J.-Q.~Li, W.~Scheid, Phys.~Rev.~C~{\bf 76}, 
044606 (2007).

\bibitem{r6}
R.S.~Naik {\it et al.}, Phys.~Rev.~C~{\bf 76}, 054604 (2007).

\bibitem{r7}
G.G.~Adamian, N.V.~Antonenko, W.~Scheid, Phys.~Rev.~C~{\bf 69}, 014607 (2004).

\bibitem{r8}
V.I.~Zagrebaev, Y.~Aritomo, M.G.~Itkis, Y.T.~Oganessian, M.~Ohta, 
Phys.~Rev.~C~{\bf 65}, 014607 (2001).

\bibitem{r9}
V.I.~Zagrebaev, W.~Greiner, Nucl.~Phys.~{\bf A944}, 257 (2015).

\bibitem{r10}
W.~Loveland, Eur.~J.~Phys.~{\bf A51}, 120 (2015).

\bibitem{r11}
W.~Loveland, J.~Phys.~Conf.~Series {\bf 420}, 012004 (2013).

\bibitem{r12}
http://nrv.jinr.ru/nrv/

\bibitem{r13}
V.G.~Sargsyan (private communication).

\bibitem{r14}
H.~Lu, D.~Boilley, EPJ Web of Conferences {\bf 62}, 03002 (2013).

\bibitem{r15}
R.~Yanez {\it et al.}, Phys.~Rev.~Lett.~{\bf 112}, 152702 (2014).

\bibitem{r16}
M.G.~Itkis {\it et al.}, AIP Conf.~Proc.~{\bf 853}, 231 (2006).

\bibitem{r17}
H.A.~Kramers, Physica {\bf 7}, 284 (1940).

\bibitem{r18}
A.N.~Andreyev {\it et al.}, Heavy Ion Fusion: Exploring the Variety of 
Nuclear Properties (Singapore, World Scientific, 1994), 260.

\bibitem{r19}
A.V.~Afanasjev, H.~Abushara, P.~Ring, Int.~J.~Mod.~Phys.~E {\bf 21}, 
1250025 (2012).

\bibitem{r20}
G.~Bertsch, W.~Loveland, W.~Nazarewicz, P.~Talou, 
J.~Phys.~G {\bf 42}, 1 (2015).

\bibitem{r21}
M.~Kowal, P.~Jachimowicz, A.~Sobiczewski, Phys.~Rev.~C~{\bf 82}, 014303 (2010).

\bibitem{r22}
A.~Baran, M.~Kowal, P.-G.~Reinhard, L.M.~Robledo, A.~ Staszczak, M.~Warda, 
Nucl.~Phys.~{\bf A944}, 442 (2015).

\bibitem{r23}
J.C.~Pei, W.~Nazarewicz, J.A.~Sheikh, A.K.~Kerman, 
Phys.~Rev.~Lett.~{\bf 102}, 192501 (2009).

\bibitem{r24}
G.~Giardina {\it et al.}, J.~Phys.~Conf.~Series {\bf 282}, 012006 (2011).

\bibitem{r25}
S.~Hofmann {\it et al.}, Eur.~J.~Phys.~{\bf A52}, 116 (2016).

\bibitem{r26}
V.~Zagrebaev, W.~Greiner, Phys.~Rev.~C 78, 034610 (2008).

\bibitem{r27}
G.N.~Knyazheva {\it et al.}, Phys.~Rev.~C~{\bf 75}, 064602 (2007).

\bibitem{r28}
E.M.~Kozulin {\it et al.}, Phys.~Rev.~C~{\bf 90} 054608 (2014).

\bibitem{r29}
R.~du Rietz {\it et al.}, Phys.~Rev.~C~{\bf 88}, 054618 (2013).

\bibitem{r30}
A.~Wakhle {\it et al.}, Phys.~Rev.~Lett.~{\bf 113}, 182502 (2014).

\bibitem{r31}
W.Q.~Shen {\it et al.}, Phys.~Rev.~C~{\bf 36}, 115 (1987).

\end{thebibliography}
\end{document}